\documentstyle [12pt]{article}
\begin{document}

\hfill {WM-96-107}

\hfill {July 1996}

\vskip 1in   \baselineskip 24pt
{

   \bigskip
   \centerline{\bf The Coleman-Weinberg Phase Transition in Extended Higgs
Models }
 \vskip .8in
 
\centerline{Marc
Sher } 
\bigskip
\centerline {\it Physics Department, College of William and
Mary, Williamsburg, VA 23187, USA}

\vskip 1in
 
{\narrower\narrower   In Coleman-Weinberg symmetry breaking, 
all dimensionful
parameters vanish and the symmetry is broken by loop corrections.  
Before
Coleman-Weinberg symmetry breaking in the Standard Model was 
experimentally
ruled out, it had already been excluded on cosmological grounds.  
In this Brief
Report, 
the cosmological analysis is carried out for Coleman-Weinberg 
models with
extended Higgs sectors, which are not experimentally ruled out, and 
general
constraints on such models are given.}

\newpage
In the Standard Model, supplemented by General Relativity, there are 
only two
dimensionful parameters:  the squares of the Higgs mass term, $\mu$, 
and the
Planck mass. 
The ratio of these two numbers must be smaller than $10^{-34}$.  In a
fundamental theory, this ratio must be explained.  A significant 
fraction of 
research in theoretical physics in the past fifteen years has focused 
on models
which can explain such a small number.

It might eventually be easier to explain this small ratio if it 
vanishes
completely.   This is somewhat similar to the assumption made in 
cosmology that
the cosmological constant vanishes; since it must be less (in Planck 
units)
than $10^{-122}$, it is generally assumed that zero will be an easier 
number to
explain.  Likewise, the $\mu^2=0$ limit could also prove easier to 
explain than
$\mu^2\sim 10^{-34}$ in Planck units.  The $\mu^2=0$ case was 
originally considered by Coleman
and E. Weinberg(CW)\cite{cw}, who showed that loop corrections will 
still break
the electroweak symmetry.

 Since a parameter has been set to zero, a
prediction can be made, and the Higgs mass can be determined in terms 
of the
gauge boson and fermion masses.  The result gives $m_h < 10$ GeV, 
and requires
$m_t < m_Z$, which are both in contradiction with experiment.  As a 
result,
Coleman-Weinberg symmetry breaking in the Standard Model is ruled out.  
It is
not ruled out, however, in models with more complicated Higgs
sectors\cite{pr}.  

Models with extended Higgs sectors have become more popular recently 
with the
realization that sphaleron effects could erase any previously 
generated baryon
asymmetry\cite{sphaleron}, and that CP violation in the standard model 
is too
small for electroweak baryogenesis\cite{cpsmall}.  It has been
shown\cite{extended} that models with extended Higgs sectors can avoid 
washing
out any previous baryon number, as well as generate a sufficient baryon
asymmetry.  
In fact, generation of a baryon asymmetry also requires a first order
electroweak phase transiton, and CW models always have a first
order transition.  This leads to greater incentive for considering CW
symmetry breaking in extended Higgs models.

Recently, several papers\cite{several,hemp} have discussed CW 
symmetry breaking
in extended Higgs models.  
In particular, a simple model of Hempfling\cite{hemp}
showed that addition of a singlet field yields a satisfactory mass 
spectrum in
the CW case. The details of the cosmological phase transition in these 
models,
however, were not considered, and this poses a potential difficulty
 for these
models.

Before experiments ruled out CW symmetry breaking in the Standard Model, 
it had
been shown\cite{pr,fs} that it was already ruled out on cosmological 
grounds;
either the phase transition never finished or it finished with far 
too much
entropy generation to be acceptable.  In this Brief Report, I will 
examine this
issue 
for the general case, and will present a bound which can be applied 
to any
specific model of electroweak CW symmetry breaking.

First, let us review the cosmology of the CW model in the Standard Model. 
Minimizing 
the potential\cite{pr}, one finds the zero-temperature potential to be
\begin{equation}
V= B\phi^4\left(\ln(\phi^2/\langle\phi\rangle^2)-{1\over 2}\right)
\end{equation}
where
\begin{equation}
B\equiv{1\over 64\pi^2\langle\phi\rangle^4}(6M^4_W+3M^4_Z-12M^4_t)
\end{equation}
and we must have $B>0$ to have spontaneous symmetry breaking. One 
immediately
sees why CW symmetry breaking is ruled out for the known top quark mass, 
since
that gives $B<0$.  However, suppose the top mass were smaller, 
say $40$ GeV;
then we would have $B>0$. The Higgs mass would then be given by
$m^2_H=8B\langle\phi\rangle^2$.

 At finite temperature, a
temperature dependent term must be added.  For values of $T>>\phi$, 
this gives
a term 
\begin{equation}
V_T=-{1\over 90}\eta \pi^2T^4+{1\over
24\langle\phi\rangle^2}\phi^2T^2(6M^2_W+3M^2_Z+6M^2_t)
\end{equation}
which causes a positive $\phi^2$ term to appear, restoring the 
symmetry at high
temperatures, and causing a minimum to occur at the origin at 
any temperature.
As a result, the Universe will start in the symmetric vacuum, and 
will have to
tunnel through a potential barrier.
 Although the above expression is only valid at
high temperature, the results of tunnelling out of the vacuum at the 
origin at
high temperature 
are not appreciably affected by using the exact expression (note
that this is not the case in determining, for example, the critical 
temperature
at which the symmetric and asymmetric vacua are 
degenerate\cite{critical}).

The tunnelling rate for this model was calculated\cite{gw} long ago, 
and it was
found 
that the transition will not occur until the temperature is 
below $10^{-7}$
times the electroweak scale.  This will result in an entropy increase 
during
the transition by a factor of more than $10^{21}$, washing out any 
baryon
asymmetry, and thus the model was ruled out.  However, 
Witten\cite{witten}
pointed out that the Universe can not cool below the QCD scale 
without chiral
symmetry 
breaking giving $\overline{\psi}\psi$ a vacuum expectation value, which
breaks the electroweak gauge symmetry (the Yukawa term,
$\overline{\psi}\psi\Phi$ turns into a linear term when chiral symmetry 
breaks,
thus destabilizing the symmetric vacuum.  The entropy generated in this
transition 
is roughly the cube of the ratio of the electroweak to QCD scales, or
$\sim 10^6$, which is marginally acceptable.  Finally, Flores and 
Sher\cite{fs}
noted that, in addition to the QCD coupling growing as the Universe 
cools, the
Yukawa coupling also grows.  This will cause the coefficient
$B$ 
in the above to change sign at low temperatures, causing a barrier 
to appear
which persists to zero temperature.  Thus, the Universe will get 
stuck in a
metastable vacuum, and if the transition occurs, far too much entropy 
would be
genrated.  Thus, they concluded that CW symmetry breaking in the Standard
Model is ruled out.
 
What happens in extended Higgs models?  The question of CW symmetry 
breaking in
extended 
Higgs models was first discussed by Gildener and 
Weinberg\cite{gild} and
reviewed 
in Ref. 2.   Consider the single Higgs case.  There, one can choose the
renormalization scale such that the tree-level potential vanishes, 
i.e. one can
choose $M_R$ such that $\lambda(M_R)=0$.  Calculating the one loop 
potential,
and 
eliminating $M_R$ by minimizing and denoting the minimum 
$\langle\phi\rangle$
gives the expression in Eq. 1.  Now consider the multi-scalar case.  
The
tree-level 
potential is given by $V_o=f_{ijkl}\phi_i\phi_j\phi_k\phi_l$ where the
$\phi_i$ label all of the scalars in the model.  As Gildener and 
Weinberg show,
one 
cannot choose the renormalization scale such that $V_o$ vanishes 
everywhere,
since 
the scale at which one self-coupling vanishes is not the same as the 
scale
at which 
another will vanish.  However, suppose one defines $\phi_i=N_i r$, 
where
$N_i$ is a unit vector and $r$ is the distance from the origin of 
field space,
then $V_o=f_{ijkl}N_iN_jN_kN_lr^4$.  Let the minimum value of $V_o$ on 
the unit
sphere occur for $N_i=n_i$.  Then, Gildener and Weinberg show that the
renormalization scale can be chosen so that $V_o$ vanishes along the 
direction
$N_i=n_i$.

The important point is that the tree level potential vanishes 
along a direction
in field space, and in any other direction will be positive (if one has
positivity 
of the potential at very large scales).  Note, the condition for this
to occur is a {\it single} condition on the $f_{ijkl}$; i.e. one can 
choose the
renormalization 
scale so that a combination of the $f_{ijkl}$ vanishes.  Now, if
the tree-level potential vanishes along the ray $\phi_i=n_ir$, then one
can calculate the potential along that ray.  The result is (minimizing 
along
that ray)
\begin{equation}
V=B r^4\left(\ln(r^2/\langle r\rangle^2)-{1\over 2}\right)
\end{equation}
where
\begin{equation}
B\equiv{1\over 64\pi^2\langle r\rangle^4}\left(3M^4_V+M^4_S-4M^4_F
\right)
\end{equation}
and where $M_V^4$, for example, refers to the sum of the fourth 
powers of all
vector boson masses. In any other direction, the tree-level potential is
positive, the
$f_{ijkl}$ 
are not small, and radiative corrections are unimportant.  The scalar
masses-squared 
are all positive, with the scalar whose mass vanishes at tree-level
(corresponding to the flat direction) getting a mass-squared of $8B
\langle
r\rangle^2$, just as in the Standard Model.

Let us give a couple of examples.  In the two-Higgs model, considering 
the
neutral directions only, we can write the doublets as
\begin{equation}
\Phi_1={r\over\sqrt{2}}\left({0\atop N_1}\right),\qquad \Phi_2=
{r\over\sqrt{2}}\left({0\atop N_2}\right)
 \end{equation}
Expressing the potential (in the massless theory--see Ref. 2 for the full
potential) in terms of these coordinates gives
\begin{equation}
V={1\over
4}r^4[\lambda_1N_1^4+\lambda_2N_2^4+(\lambda_3+\lambda_4+
\lambda_5)N^2_1N^2_2]
\end{equation}
Suppose the potential has a minimum value (which Gildener and 
Weinberg show can
be chosen to be zero by judicious choice of renormalization scale) 
on the
unit circle along the direction $N_i=n_i$.  This direction can be 
found to be
\begin{equation}
n_1={\sqrt{\lambda_2}\over \sqrt{\lambda_1}+\sqrt{\lambda_2}},\qquad n_2=
{\sqrt{\lambda_2}\over \sqrt{\lambda_1}+\sqrt{\lambda_2}},\qquad
2\sqrt{\lambda_1\lambda_2}+\lambda_3+\lambda_4+\lambda_5=0.
\end{equation}
The first two of these specify the direction from the origin to 
the minimum and
the third comes from the requirement that the potential vanish 
along this
direction (this is no more ``unnatural" that the assumption that a
renormalization scale can be found in the Standard Model such that 
$\lambda$ is
small).  The one-loop potential along that direction is then
$V=Br^4[\ln(r^2/\langle r\rangle^2)-{1\over 2}]$ where $\langle
r\rangle^2
=\langle\phi_1\rangle^2+\langle\phi_2\rangle^2= (246\ {\rm GeV})^2$ 
and
$B$, 
given in Eq. 5, includes contributions from the $W,Z,t$ and the 
five scalars
in the model.  The lightest Higgs mass-squared is $8B\langle r
\rangle^2$.  This
mass 
can exceed the experimental limit only if some of the scalar masses 
are quite
large (greater than several hundred GeV).

Another example is the model of Hempfling\cite{hemp}, with a 
singlet $S$ added.
The tree-level potential is $V_o={1\over 2}\lambda_\phi(\phi^
\dagger\phi)^2+
{1\over 2}\lambda_S(S^\dagger S)^2-\lambda_X(\phi^\dagger\phi)
(S^\dagger S)$.
Hempfling defines $\phi\equiv r\sin\beta$ and $S\equiv r\cos\beta$; 
one sees
that $r$ has the same definition as the above and $\tan\beta = 
n_2/n_1$ above.
The flat direction is given by $\tan\beta=\lambda_X/\lambda_\phi$ 
and the
condition that the tree-level potential vanish at the minimum is 
given  by
$\lambda_\phi\lambda_S=\lambda_X^2$.  The one-loop potential is then 
given by
$V=Br^4[\ln(r^2/\langle r\rangle^2)-{1\over 2}]$
where
 $B$ is also given in Eq.5.   Note that a major difference is that 
$\langle
r\rangle$ is not, as in the previous example, constrained to be 
$246$ GeV, but
could be much larger, since the singlet vacuum expectation value is
unconstrained.  This means, for example, that the contributions 
to $B$ from the
$W$, $Z$ and top quark are smaller than in the standard model by 
a factor of
$\left({246\ {\rm GeV}\over r}\right)^4$ (this can be seen in the 
$\sin^4\beta$
factor in Eq. 4 of Hempfling\cite{hemp}).  As a result, $B$ 
can be positive
without requiring enormous scalar masses.
 Again, the lightest
Higgs 
mass-squared is $8B\langle r\rangle^2$, in agreement with 
Hempfling.  Note
that since $\langle r\rangle$ can be quite large,  this mass can 
easily satisfy
the experimental bounds.

We 
now include the effects of high temperature.  Note that the potential 
is flat
along one direction, and large and positive in other directions.  
We can thus
effectively consider the problem to  be a one-dimensional problem 
along the
flat direction.  In the high temperature limit, the 
temperature-dependent term
can be written as 
\begin{equation}
V_T={1\over 24} r^2 T^2 M_2
\end{equation}
where
\begin{equation}
M_2={3M^2_V+M^2_S+2M^2_F\over \langle r\rangle^2}
\end{equation}
and 
$M_V^2$ ($M^2_S,\ M^2_F)$ stands for the sum of the squares of the 
masses of 
all of the vector bosons (scalars, fermions) in the model.  Note 
that $M_2$ is
always positive, thus CW models always have symmetry restoration 
at high
temperatures (unlike non-CW models, which, in some cases, can have 
symmetry
anti-restoration\cite{pr}).  As discussed above, we have verified 
in several
sample cases that the high-temperature limit is satisfactory for this
calculation, to within the accuracy stated.  

The full potential is thus relatively simple.  It has two parameters, 
$B$ and
$M_2$, and is a function only of the radial co-ordinate, $r$.  We 
can now
examine the phase transition as a function of these two 
parameters\cite{bert}.
The procedure is standard and straightforward--the reader is 
referred to Ref. 2
for a review.  The transition temperature is found by comparing the 
bubble
nucleation 
rate with the expansion rate (the O(3) symmetric action is always the
lowest here), and the reheating temperature is found by equating 
the energy
densities just prior to and just after the transition.  The cube 
of the ratio
(times a factor of O(1) involving the number of states) gives 
the increase
in entropy generated by the transition.

The ratio of the final to initial entropy is plotted in Fig. 1 
for various
values of $B$ and $M_2$.  Recall that the observed baryon number 
to entropy
ratio today is approximately $10^{-9}$, and in many models this 
requires very
efficient baryogenesis.  Should the entropy increase in the 
transition be a
factor 
of 10 or 100, then the transition will not play a significant role 
(it is
unlikely that a specific model will be able to predict the baryon 
number to
entropy ratio more accurately than an order of magnitude).  Should 
the entropy
increase be a factor of $10^4-10^6$, then an extremely efficient 
mechanism for
baryogenesis must be found.  The entropy increase cannot be much 
more than
$10^6$, since the transition temperature will be near or below the 
QCD scale,
and the transition will be driven by QCD condensate 
formation\cite{rvalue}.

For regions of parameter-space in which the transition is driven 
by QCD, the
entropy 
increase of at least $10^6$ requires an initial baryon-number 
to entropy
ratio close to $10^{-3}$.  It is difficult, but perhaps not 
impossible, to
construct a baryogenesis mechanism which can generate such a 
large ratio. 
However, such regions of parameter-space have another, more 
serious, problem. 
 
As the Universe cools in the symmetric phase, all of the quarks, 
vector bosons
and Higgs bosons are massless.  At what temperature will 
$\overline{\psi}\psi$
condense?  The QCD beta function is flatter than in the standard 
model, since
all six quarks contribute to very low scales---this will lower 
the chiral
symmetry 
breaking transition temperature by roughly (very roughly) a 
factor of 3. 
However 
another factor will increase the transition temperature.  The 
top quarks
will not only attract each other via QCD, but also via Higgs 
exchange (recall
that the Higgs boson is massless).  As the Universe cools, the 
top quark Yukawa
coupling 
will rapidly reach its fixed point, which (at one-loop) is 
${4\over 3}$
times the QCD coupling.  This additional attraction will break 
the $SU(6)$
symmetry and cause top quark condensation to occur at a higher 
temperature than
that of the other quarks.  The transition temperature will thus 
be somewhat
similar to the chiral symmetry breaking temperature (roughly 
$200$ MeV).  Note
that the idea of top quark condensation driving the electroweak 
symmetry
breaking has been of great interest\cite{hill}, but here the
condensation drives electroweak symmetry breaking below the GeV 
scale!

The 
serious problem occurs because of the growth of the top quark 
Yukawa coupling
as the Universe cools in the symmetric phase.  How does this 
affect the
potential?  
As discussed in detail in Ref. 2, the potential we have 
considered is
{\it not} ``renormalization-group improved", and neglected terms are
approximately of
$O({g^2\over 4\pi}\ln({Q_1\over Q_2}))$ times those included, where 
$g$ is a
coupling and
$Q_1$ and
$Q_2$ are the largest and smallest scales in the region of interest.  
Since we
are only considering values of $r$ between the QCD and electroweak 
scales, the
logarithm is not too large, and the effects are only relevant for 
the top quark
Yukawa coupling.  In this case, it is sufficient to ``run" the 
top quark
Yukawa coupling (and QCD coupling, which enters in the beta function 
for the
Yukawa coupling) in the expression for $B$ (this has been verified 
using the
full expression).  Thus, as the top quark Yukawa coupling 
grows, $B$ will
eventually change sign.  The potential near the origin, instead 
of being
negative at zero temperature, will be positive, leading to a 
barrier at zero
temperature.  Even when chiral symmetry breaks, this barrier 
is likely to
persist, and the Universe will get stuck in a metastable state; 
significantly
increasing the entropy generation further.  Although pertubative 
techniques are
questionable at these low scales, it does appear that such models, 
already
endangered by the entropy production required, are in further 
jeopardy, and
likely ruled out.

The 
results of this Brief Report can be summarized as follows.  
For any CW model,
one finds the sum of the squares of all scalar vacuum expection 
values, calling
it $\langle r\rangle^2$.  Plugging that and the masses of all of the 
particles
in the model into Eqs. 5 and 10, one finds
$B$ and
$M_2$. 
Looking at Figure 1, the entropy generation can be determined.  
If it is
less 
than $10^2$, the effects of the transition are small; if it is 
greater than
$10^6$, 
in the QCD region, the transition is apparently fatal.  
In between, very
efficient baryogenesis is required.  Let us illustrate this 
procedure with two
simple examples.  In the Standard Model with a 40 GeV top quark, 
we have 
$\langle r\rangle=246$ GeV and thus $M_2=1.22$ and $64\pi^2B=0.12$, 
and from
Figure 1 
we see that the model is excluded, in agreement with previous results.  
In the Hempfling model, in the limit in which $\langle r
\rangle>>246$ GeV, the
only relevant scalar mass is given by
$M_S^2=6\lambda_S\langle r\rangle^2$ and there is a gauge boson 
of mass
$g_X\langle r\rangle$.  This gives a value of $M_2=6\lambda_S+
3g_X^2$ and
$64\pi^2B=36\lambda_S^2+3g_X^4$.  For $\alpha_X=.01$, we find that 
the entropy
generation 
is less than $10^2$ for $\lambda_S>.15$, is between $10^2$ and 
$10^6$
for $.10<\lambda_S<.15$ and is greater than $10^6$ (i.e. the model 
is excluded)
for 
$\lambda_S<.10$.  A similar procedure can be applied to any 
Coleman-Weinberg
electroweak model.

I thank Carl Carlson for useful discussions.  This work was 
supported by the
National Science Foundation.

\def\prd#1#2#3{{\rm Phys.~Rev.~}{\bf D#1}, #2 (19#3)}
\def\plb#1#2#3{{\rm Phys.~Lett.~}{\bf B#1}, #2 (19#3)}
\def\npb#1#2#3{{\rm Nucl.~Phys.~}{\bf B#1}, #2 (19#3)}
\def\prl#1#2#3{{\rm Phys.~Rev.~Lett.~}{\bf #1}, #2 (19#3)}

\bibliographystyle{unsrt}

\begin{thebibliography}{99}
\bibitem{bert}
Our approach is similar to that of O. Bertolami, \plb{176}{416}{86}.  
In that
work, however, the expression for the high-temperature limit is 
incorrect, the
thin wall approximation (which may not be valid) is used, and the 
author only
considers the condition in which the transition is driven by QCD.
\bibitem{rvalue}
In making this statement, we assumed that $r$ was $246$ GeV at 
the minimum.  In
models with singlets, it could be closer to the TeV scale, in which 
case the
entropy generation which occurs if the transition is driven by QCD 
condensates
is more like $10^8$ than $10^6$.  As one can see from Fig. 1, this 
makes very
little difference in the result.
\end{thebibliography}

\begin{figure}

\vglue 6in  
%\hskip 1.00in {\special{picture penguins scaled 1000}} \hfil
\vglue 0.1in

\caption{For any electroweak CW model, the values of $M_2$ and $B$, 
as defined
in the text, can be determined.  Here is plotted the entropy 
generated in the
phase transition as a function of those parameters.  In the region 
below the
bottom 
line, the transition is broken by $\overline{t}t$ condensation; since the
entropy 
generated is at least $10^6$, and a metastable vacuum is likely to form,
this region is apparently excluded.  For a given $M_2$, the 
uncertainty in
$64\pi^2B$ 
due to the high-temperature approximation and due to uncertainties in
the expansion rate is approximately $0.2$.  For $M_2$ below $0.5$, the
uncertainties 
are larger, however the value of $M_2$ exceeds $0.5$ in all models
that have appeared to date.}
\end{figure}

\end{document}